\documentclass[twocolumn, longbibliography, prl, amsmath, amssymb,superscriptaddress]{revtex4-1}
\usepackage{amsfonts}
\usepackage{graphicx}
\usepackage[colorlinks,linkcolor=blue,citecolor=blue,urlcolor=blue]{hyperref}
\usepackage{float}
\usepackage[normalem]{ulem}

\begin{document}
\title{Non-Markovian Stochastic Resonance of Light in a Microcavity}

\author{K. J. H. Peters}
\affiliation {Center for Nanophotonics, AMOLF, Science Park 104, 1098 XG Amsterdam, The Netherlands}

\author{Z. Geng}
\affiliation {Center for Nanophotonics, AMOLF, Science Park 104, 1098 XG Amsterdam, The Netherlands}

\author{K. Malmir}
\affiliation {Department of Materials, University of Oxford, Parks Road, Oxford OX1 3PH, UK}

\author{J. M. Smith}
\affiliation {Department of Materials, University of Oxford, Parks Road, Oxford OX1 3PH, UK}

\author{S. R. K. Rodriguez}  \email{s.rodriguez@amolf.nl}
\affiliation {Center for Nanophotonics, AMOLF, Science Park 104, 1098 XG Amsterdam, The Netherlands}

\begin{abstract}
We report the first observation of non-Markovian stochastic resonance, i.e., noise-assisted amplification of a periodic signal in a system with memory. Our system is an oil-filled optical microcavity which, driven by a continuous wave laser, has memory in its nonlinear optical response.  By modulating the cavity length while adding noise to the driving laser, we observe a peak in the transmitted signal-to-noise ratio as a function of the noise variance. Our experimental observations are reproduced by numerical simulations, which further reveal that the stochastic resonance bandwidth is enlarged by the memory time of the nonlinearity. This frequency range available for noise-assisted amplification is $10^8$ times greater in our oil-filled cavity than in a Kerr nonlinear cavity. Our results pave the way for exploring the interplay of nonlinearity, noise, and memory, in oil-filled cavity arrays, where non-Markovian dynamics could enhance noise-assisted transport and synchronization effects.
\end{abstract}
\date{\today}
\maketitle

Stochastic resonance (SR) is a phenomenon wherein an optimum amount of noise amplifies the response of a nonlinear system to a periodic signal~\cite{Gammaitoni_RMP1998}. The essence of SR can be recognized in the behavior of a Brownian particle in a double well potential (DWP). As explained by Kramers~\cite{Kramers}, a Brownian particle escapes from a potential well in a characteristic time $\tau_{\mathrm{esc}} \propto \textrm{exp}(\Delta V / k_BT) $,  with $\Delta V$ the energy barrier  and  $k_BT$ the energy of thermal fluctuations. If the potential is modulated with period $T_\mathrm{mod}=2\tau_{\mathrm{esc}}$, the particle's Brownian motion can synchronize with the modulation. This results in two barrier crossings per period --- stochastic resonance.

SR was first proposed as an explanation for recurrent ice ages~\cite{Benzi_JPhysA1981, nicolis1993}. Since then, SR has been extensively observed in physics~\cite{Fauve83, McNamara_PRL1988, hibbs1995, Lee03, Abbaspour_PRL2014, Abbaspour_PRB2015, Venstra13, Monifi_NatPhot2016, Ricci17, Chowdhury17, Wagner19},  chemistry~\cite{leonard1994, guderian1996, forster1996, hohmann1996}, biology~\cite{Douglass_Nat1993, bezrukov1995, jaramillo1998, Russell_Nat1999, Mori02}, ecology~\cite{Blarer1999}, psychophysics~\cite{collins1996,simonotto1997,ward2002}, climate science~\cite{Ganopolski_PRL2002}, finance~\cite{xiao2002, krawiecki2003} and social science~\cite{wallace1997}. Applications of SR to imaging~\cite{Marks2002, rallabandi2008, dylov2010} and mehanical engineering~\cite{Niaoqing, li2013} have also emerged. To date, all observations of SR have been described within the Markov approximation neglecting memory effects. While non-Markovian dynamics have been experimentally observed in various systems~\cite{Hanggi85, Liebovitch87, Mercik01, Houlihan04, Piilo11, madsen2011, hoeppe2012,  lu2020} and are theoretically expected to modify SR~\cite{Neiman96, Goychuk03, Prager03}, non-Markovian SR has not been experimentally reported.

In this Letter, we report the first observation of non-Markovian stochastic resonance.  We investigate an oil-filled optical microcavity with memory in its nonlinear optical response. In contrast to the widely studied class of non-Markovian systems introduced by Mori~\cite{Mori} and described by generalized Langevin equations~\cite{Hanggi78}, the memory of our cavity is unrelated to its dissipation. Moreover, non-Markovian dynamics emerge even for Gaussian white noise. Here, we  evidence SR  in this non-Markovian regime by driving our oil-filled cavity with a continuous wave laser to which we add a controlled amount of Gaussian white noise.  By periodically modulating the cavity length we imprint a signal on the laser, and we measure the transmitted signal-to-noise ratio ($SNR$) as a function of the added noise variance. SR manifests as a peak in the $SNR$ at a certain noise variance. Our experimental observations are reproduced by numerical simulations based on a recently developed model accounting for the memory time of the nonlinearity~\cite{Geng_PRL2020}. Simulations show how the memory time of the nonlinearity enlarges the SR bandwidth, making it 8 orders of magnitude greater for our oil-filled cavity than for conventional Kerr nonlinear cavities. This memory-enhanced robustness of SR against changes in signal frequency opens new perspectives for harvesting energy from fluctuations across an unprecedentedly large bandwidth.

\begin{figure}[!b]
	\includegraphics[width=\columnwidth]{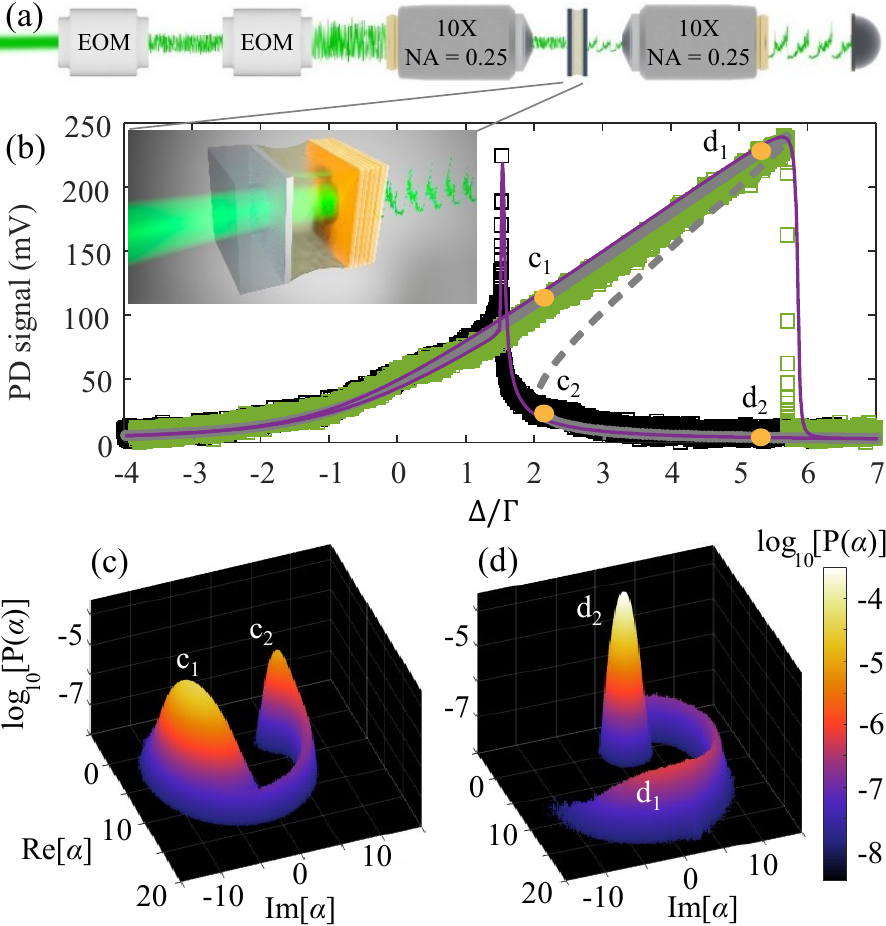}
	\caption{\label{fig:1}(a) Illustration of an oil-filled optical microcavity driven by a noisy laser. Electro-optic modulators (EOMs) add amplitude and phase noise to a $532$ nm continuous wave laser. Light is coupled into and out-of the microcavity using microscope objectives. The cavity length is controlled with a piezoelectric actuator, and the transmitted power is measured with a photodetector.    (b) Measurements of the transmitted power when slowly closing and opening the cavity, shown as  black and green curves respectively. The cavity length was converted to $\Delta/\Gamma$ (see Eq.~\ref{eq:IDE})  using the laser frequency and the cavity resonance linewidth at low power. Solid and dashed gray curves are stable and unstable steady-states, respectively. Purple curves are dynamic hysteresis simulations using Eq.~\ref{eq:IDE} with $D=0$. (c)-(d) PDFs in the complex $\alpha$-plane calculated by evolving Eq.~\ref{eq:IDE} with different realizations of the noise. $\Delta/\Gamma=2.007$ in  (c)  and $\Delta/\Gamma=5.58$ in (d). The correspondence between the peaks in the PDFs in (c)-(d) and the states in (b) is indicated by the matching symbols in the two panels. }
\end{figure}

Figure~\ref{fig:1}(a) illustrates our experimental setup, comprising a tunable microcavity filled with macadamia oil. The cavity is made by a planar and a concave mirror. The planar mirror is a 60 nm thick silver layer on a glass substrate. The concave mirror ($7$ $\mu$m diameter, $12$ $\mu$m radius of curvature) is  made by milling a glass substrate with a focused ion beam~\cite{Trichet15}, and subsequently coating it with a distributed Bragg reflector (DBR). The DBR has a peak reflectance of $99.9$\% at $530$ nm. We use a six degree-of-freedom piezoelectric actuator to align and position the concave mirror, and a single high-frequency piezoelectric actuator to periodically displace the planar mirror. We can probe a single optical mode when displacing the planar mirror across several nanometers because the micron-scale dimensions of the concave mirror strongly confine the optical modes. The cavity is driven by a $532$ nm single-mode continuous wave laser, which heats the oil and causes its refractive index to decrease. A similar  intensity-dependent refractive index underlies the observation of SR in Kerr nonlinear cavities at cryogenic temperatures~\cite{Abbaspour_PRL2014}, but with two important differences.  Our oil-filled cavity has a strong nonlinearity at room temperature, and non-instantaneous thermal relaxation endows our system with memory, thereby enabling us to explore non-Markovian regimes. Our excitation and collection objectives both have $10\times$ magnification and a numerical aperture of $0.25$. In all experiments here presented, we drive the $15^\text{th}$ longitudinal cavity mode with a laser power of $7.8$ mW at the excitation objective. The transmitted light is measured with a photodetector and an oscilloscope.

In a frame rotating at the laser frequency $\omega$, the light field $\alpha$ in our single-mode oil-filled cavity satisfies the following stochastic integro-differential equation:
\begin{widetext}
    \begin{equation}\label{eq:IDE}
    	i \dot{\alpha}(t) = \left[-\Delta- i\frac{\Gamma}{2} + U \int_0^t ds\,K(t-s)\left(|\alpha(s)|^2-1\right)\right]\alpha(t) + i \sqrt{\kappa_L}F + \frac{D}{\sqrt{2}} \left[\xi_1(t) + i \xi_2(t) \right].
    \end{equation}
\end{widetext}

\noindent  $\Delta=\omega-\omega_0$ is the laser-cavity detuning, with $\omega_0$ the resonance frequency. $\Gamma=\gamma+\kappa_L+\kappa_R$ is the total loss rate, with $\gamma$ the absorption rate and  $\kappa_{L,R}$ the input-output rates through the left and right mirror.  $U$ is the strength of the cubic nonlinearity, with non-instantaneous character captured by the  time-integral in Eq.~\ref{eq:IDE}. The integral contains the  memory kernel $K(t)=\mathrm{exp}(-t/\tau)/\tau$ with  $\tau$ the thermal relaxation time. This form of the kernel ensures that steady-states ($\dot{\alpha}=0$ in Eq.~\ref{eq:IDE})   are the same as for an instantaneous nonlinearity~\cite{Geng_PRL2020}. $F$ is the laser amplitude.  $\xi_1(t)$ and $\xi_2(t)$ provide Gaussian white noise in the laser amplitude and phase; the combined variance of these stochastic processes is $D^2$. Note that for $\tau \rightarrow 0$ and $D=\sqrt{\Gamma/2}$, Eq.~\ref{eq:IDE}  describes a Kerr nonlinear cavity influenced by quantum fluctuations within the so-called truncated Wigner approximation~\cite{CarusottoRMP}. In Supplemental Material we provide further details about our calculations based on the xSPDE toolbox~\cite{xSPDE} for \textsc{Matlab}, and the values of the model parameters we used~\cite{supp}.

For vanishing noise ($D=0$) and a sufficiently large $F$ making the mean-field interaction energy larger than the losses (i.e., $U |\alpha|^2 \gtrsim  \Gamma$), Eq.~\ref{eq:IDE} predicts optical bistability: two steady-states with different intracavity intensity $|\alpha|^2$ at a single driving condition. To observe bistability, we measure the transmitted laser power at constant $F$ while opening and closing the cavity. The cavity length maps to $\Delta/\Gamma$, with $\Gamma$ the linewidth at low power. For example,  Fig.~\ref{fig:1}(b) shows transmission measurements where bistability occurs for $2 \lesssim \Delta/\Gamma \lesssim 5.5$. The measurements also display an overshoot at $\Delta/\Gamma\sim 1.5$, due to the oil's non-instantaneous thermal relaxation~\cite{Geng_PRL2020}. Steady-state calculations assuming instantaneous nonlinearity [gray curves in Fig.~\ref{fig:1}(b)]  do not reproduce this overshoot. In contrast, dynamic simulations [purple curves in Fig.~\ref{fig:1}(b)] based on Eq.~\ref{eq:IDE} including memory effects reproduce all our observations, including the overshoot. The width of the overshoot is determined by $T_\mathrm{mod}/\tau$. In  Supplemental Material we show that  $\tau=10$ $\mu$s in our oil-filled cavity~\cite{supp}.

\begin{figure}[!t]
	\includegraphics[width=\columnwidth]{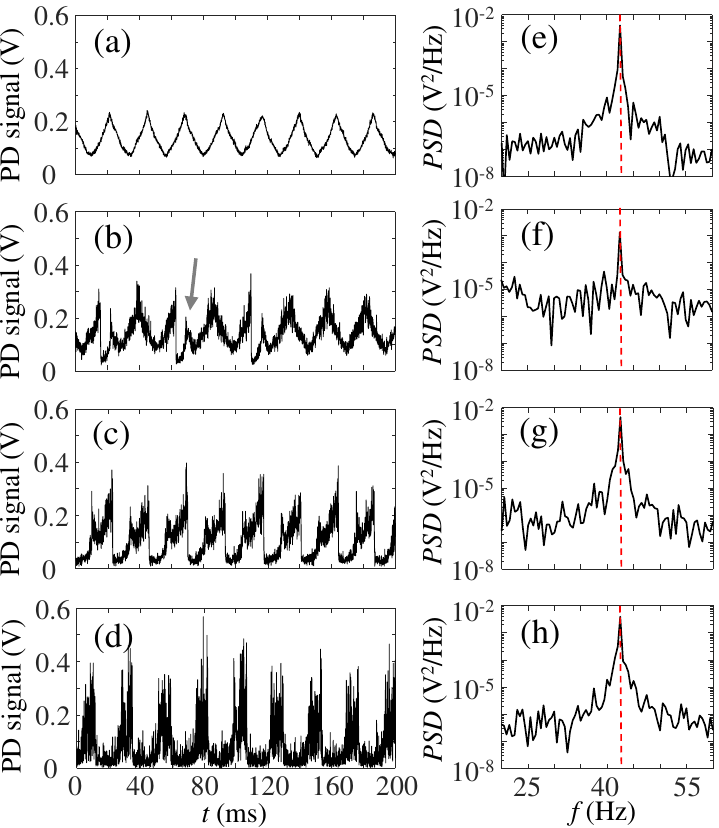}
	\caption{\label{fig:2} (a)-(d) Transmission measurements as a function of time for increasing peak-to-peak voltage $V_\mathrm{pp}$ supplied to the modulators adding noise. From top to bottom: $V_\mathrm{pp}=1$ V, $V_\mathrm{pp}=50$ V, $V_\mathrm{pp}=70$ V, $V_\mathrm{pp}=200$ V. The arrow in (b) indicates the overshoot after a switching event. (e)-(h), Power spectral densities obtained by Fourier-transforming the corresponding time traces in (a)-(d). Dashed red lines indicate the modulation frequency $f_{\textrm{mod}}$ of the cavity length.}
\end{figure}

For $D \neq 0$,  light in a bistable optical cavity behaves like a Brownian particle in a DWP. For constant $F$ and $\Delta/\Gamma$ yielding bistability, $\alpha$ randomly switches between states at the Kramers rate $\tau_\mathrm{esc}^{-1} \propto \textrm{exp}(D)$. This switching behavior can be observed in  stochastic trajectories of $|\alpha|^2$ shown in Supplemental Material~\cite{supp}. Based on multiple trajectories $\alpha(t)$, we calculate a probability density function (PDF)  in the complex $\alpha$-plane. Figures~\ref{fig:1}(c,d) show this PDF for two values of $\Delta/\Gamma$. Notice the bimodal distributions reminiscent of a Brownian particle in a DWP.
The matching symbols in Fig.~\ref{fig:1}(b) and Fig.~\ref{fig:1}(c) indicate the correspondence between  high \& low intensity states of the bistability and  the peaks in the PDF. It follows that by modulating $\Delta/\Gamma$ and for the right $D$, SR can emerge in the transmission of light through our microcavity.

To demonstrate  SR, we measure the transmitted laser signal while modulating the cavity length.  The cavity acts as a transducer, converting a mirror displacement to an optical signal.  We set a subthreshold modulation amplitude of 12 nm, which is insufficient for deterministic switching between states given our laser power. Using the setup  illustrated in Fig.~\ref{fig:1}(a), we add uncorrelated amplitude and phase noise by passing the laser through electro-optic modulators connected to different waveform generators. In Supplemental Material we show that the noise power is approximately constant up to 200 kHz~\cite{supp}, well-above our cavity modulation frequency $f_{\textrm{mod}}=42.5$Hz and above the thermal relaxation rate $\tau^{-1}=100$ kHz. Effectively, this corresponds to white noise. We also show that the standard deviation of the noise $D$ is proportional to the peak-to-peak voltage $V_\mathrm{pp}$ in the waveform generators~\cite{supp}.

Figures~\ref{fig:2}(a)-(d) and ~\ref{fig:2}(e)-(h) show measurements of the transmitted signal and the corresponding power spectral density ($PSD$), respectively, for different $V_\mathrm{pp}$.  For the smallest $V_\mathrm{pp}$,  the  transmitted power is modulated by the changing cavity length while noise plays a secondary role [Fig.~\ref{fig:2}(a)]. The corresponding $PSD$ [Fig.~\ref{fig:2}(e)] shows a  peak at $f_{\textrm{mod}}$. For increased $V_\mathrm{pp}$, noise makes the transmitted intensity switch between high and low values sometimes. These switches occur close to the maximum modulation amplitude [Fig.~\ref{fig:2}(b)]. In the corresponding $PSD$ [Fig.~\ref{fig:2}(f)], this manifests as a greater noise floor and a reduced peak at $f_{\textrm{mod}}$.  For greater $V_\mathrm{pp}$, the switching synchronizes with the modulation [Fig.~\ref{fig:2}(c)] and the peak in the $PSD$ grows [Fig.~\ref{fig:2}(g)]. Further increasing $V_\mathrm{pp}$ makes the transmitted signal  more noisy  [Fig.~\ref{fig:2}(d)]. The noise floor and the peak at $f_{\textrm{mod}}$ barely change [Fig.~\ref{fig:2}(h)], but peaks at higher harmonics substantially decrease (see Supplemental Material~\cite{supp}).   Overall, these measurements demonstrate the signature of  SR: the signal power is a non-monotonic function of the noise strength.

\begin{figure}[!t]
	\includegraphics[width=\columnwidth]{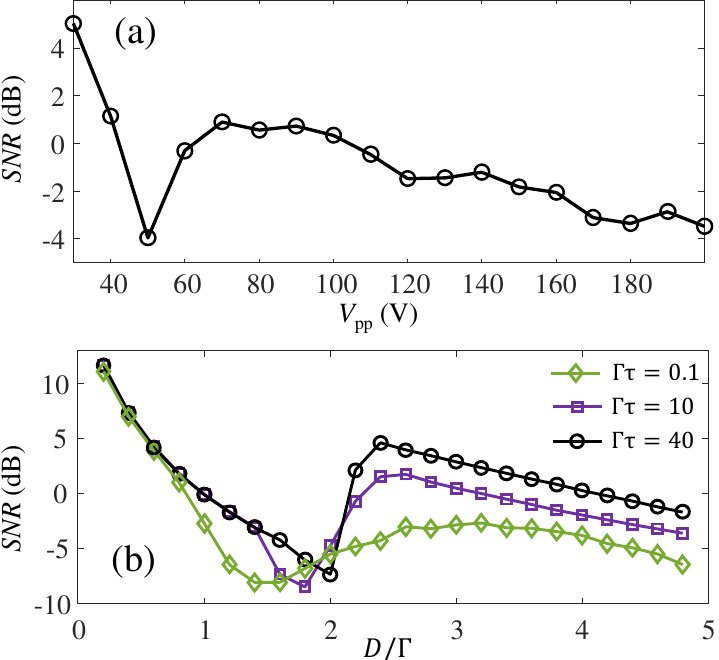}
	\caption{\label{fig:3} (a) Measured signal-to-noise ratio $SNR$ as function of the peak-to-peak voltage $V_\mathrm{pp}$ supplied to the modulators adding noise. Experimental conditions are as in Fig.~\ref{fig:2}. Each data point is an average of 10 measurements of 2 seconds ($\sim800$ modulation cycles in total). (b) Calculated $SNR$ as a function of the standard deviation $D$ of the noise (divided by the total loss rate $\Gamma$) for three values of the thermal relaxation time $\tau$. The ratio $T_\mathrm{mod}/\tau=10^3$ is kept constant. Model parameters (see Supplemental Material~\cite{supp}): $F/F_c=2.73$, $\Delta_\mathrm{min}=2.08\Gamma$, $\Delta_\mathrm{max}=5.58\Gamma$.}
\end{figure}

Next, we analyze the measured transmitted signal-to-noise ratio. We define the $SNR$ as the power in the first six harmonics including the fundamental $f_{\textrm{mod}}$, relative to the noise floor. Figure~\ref{fig:3}(a) shows the $SNR$ versus $V_\mathrm{pp}$ over an extended range. We can distinguish three regimes. For small $V_\mathrm{pp}$, the $SNR$ decreases with increasing $V_\mathrm{pp}$. This is due to an increasing number of random switching events and a growing  noise floor. For intermediate $V_\mathrm{pp}$, the $SNR$ increases with $V_\mathrm{pp}$. Here switching events increasingly synchronize with the modulation.  Finally, for large $V_\mathrm{pp}$ the $SNR$ decreases with $V_\mathrm{pp}$. This effect can be observed in the reduced amplitude of the signal harmonics as shown in Supplemental Material~\cite{supp}. In combination, these three regimes result in a SR peak in the $SNR$ around $V_\mathrm{pp}=70$ V.

We further validate our model by reproducing our $SNR$ measurements. This will enable us to confidently simulate the effect of the memory time $\tau$ on SR  over an extended range,  which is not easily done experimentally. Figure~\ref{fig:3}(b) shows results of simulations corresponding to the measurements in Fig.~\ref{fig:3}(a). We consider three values of $\tau$,  keeping the ratio  $T_\mathrm{mod}/\tau$ constant since it determines the detuning range for which bistability occurs~\cite{Geng_PRL2020}. For  $\tau>\Gamma^{-1}$, we observe a $SNR$ peak growing with $\tau$ and remaining at approximately constant $D/\Gamma$. This behavior is due to two competing effects. On one hand, increasing $T_\mathrm{mod}$ makes the $SNR$ peak grow and shift to smaller $D$; the same is true in standard Markovian SR. On the other hand, increasing $\tau$ slightly reduces the $SNR$ peak and shifts it to larger $D$; this effect is unique to our non-Markovian system. We attribute the peak $SNR$ reduction to the overshoot following each switching [see Fig.~\ref{fig:2}(b)], which makes the signal further deviate from a pure sinusoidal. The shift  to larger $D$  is associated with non-exponential distributions of residence times in the metastable states of the cavity~\cite{Geng_PRL2020}, which is the hallmark of non-Markovian dynamics. Essentially, thermal relaxation imposes a high-frequency cut-off for switching events at $\tau^{-1}$. This results in SR at larger input noise variance than in the Markovian case.

\begin{figure}[!t]
	\includegraphics[width=\columnwidth]{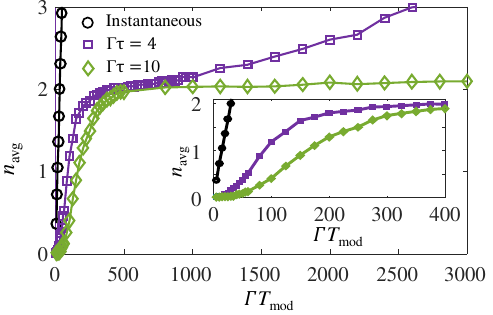}
	\caption{\label{fig:4}  Calculated average number of switches per cycle $n_\mathrm{avg}$ as a function of the modulation period $T_\mathrm{mod}$ (times the total loss rate $\Gamma$), for fixed $D=5\Gamma$. The average is computed over $1200$ modulation cycles. Inset: Zoom into fast modulations. Model parameters are as in Fig.~\ref{fig:3}.}
\end{figure}

In Fig.~\ref{fig:4} we calculate the average number of switches per cycle $n_\mathrm{avg}$ as a function of $T_\mathrm{mod}$, for different values of $\tau$. We consider a fixed noisy environment (constant $D\neq 0$), and calculate $n_\mathrm{avg}$ as explained in Supplemental Material~\cite{supp}. Recall that SR corresponds to $n_\mathrm{avg}\approx2$. For an instantaneous nonlinearity ($\tau \rightarrow 0$), the black circles in Fig.~\ref{fig:4} show how $n_\mathrm{avg}$ simply increases linearly with $T_\mathrm{mod}$. In our system, this Markovian limit is reached  for $\tau \ll \Gamma^{-1}$. More interesting behavior arises in the non-Markovian regime where $\tau > \Gamma^{-1}$. Taking  $\Gamma \tau = 4$ and $\Gamma \tau = 10$ as examples, Fig.~\ref{fig:4} shows the emergence of a `plateau' at the SR condition $n_\mathrm{avg}\approx2$. This plateau represents an enlargement of the signal frequency range in which SR can be achieved, i.e., the SR bandwidth. Thus, thanks to the slow nonlinearity, SR becomes robust to changes in signal frequency. In Supplemental Material we show that the SR bandwidth increases linearly with $\tau$~\cite{supp}.   By extrapolating the SR bandwidth calculated as a function of $\tau$ up to our experimental value $\tau = 10$  $\mu\textrm{s}$, and considering  $\Gamma^{-1} =4$ ps in our experiments, we deduce that the SR bandwidth is enlarged by a factor of $10^8$ in our oil-filled cavity relative to a Kerr nonlinear cavity.

The enlarged SR bandwidth in our non-Markovian system is associated with the suppression of fast switching events by the non-instantaneous nonlinearity. Indeed, for fast modulations satisfying $ \Gamma T_\mathrm{mod} \lesssim 100$, bistability disappears because the nonlinearity does not have time to build up~\cite{Geng_PRL2020}. This high-frequency regime is characterized by $n_\mathrm{avg}<2$, as shown in the inset of Fig.~\ref{fig:4}. In the opposite regime of slow modulations,  $n_\mathrm{avg}$ increases linearly with $T_\mathrm{mod}$ as in the Markovian case, albeit with a smaller slope. The smaller slope further enlarges the SR bandwidth.

In summary, we demonstrated noise-assisted transmission of light through an optical microcavity with non-instantaneous nonlinearity --- non-Markovian SR. Through numerical simulations we found that the frequency range available for SR grows with the memory time of the nonlinearity. Extensions of our work could investigate the effects of colored or non-Gaussian noise, which were previously shown to enhance Markovian SR in certain regimes~\cite{Nozaki99, Fuentes01,  Castro01}. Our findings can have important applications to nonlinear energy harvesting~\cite{Cottone09}. Typically, energy harvesting is limited to a narrow frequency bandwidth by the (instantaneous) response of the device. In contrast, a device with   non-instantaneous  nonlinearity could be used to extract energy from environmental fluctuations across an unprecedentedly large bandwidth.  Our results also provide a plausible explanation for the occurrence of various noise-assisted processes at different frequencies within a single noisy environment, as observed in Nature.  Beyond single-resonator physics,  our  work paves the way  to study various noise-assisted processes (e.g. noise-assisted transport~\cite{Plenio08, Caruso_JChemPhys_2009, Caruso15, Ramirez20}) in nonlinear and non-Markovian cavity arrays. Such arrays could also be used to explore how complex networks can be made resilient against breakdown of synchronization thanks to non-Markovian effects~\cite{Lin20}.  \\

\section*{Acknowledgments}
\noindent This work is part of the research programme of the Netherlands Organisation for Scientific Research (NWO). We thank Ewold Verhagen for critical comments on our manuscript, Allard Mosk, Femius Koenderink, and Sanli Faez for stimulating discussions, and Ricardo Struik and Niels Commandeur for technical support. S.R.K.R. acknowledges a NWO Veni grant with file number 016.Veni.189.039.

%\bibliography{bibfile}


\renewcommand{\thefigure}{S\arabic{figure}}
\setcounter{figure}{0}
\clearpage
\newpage

\section{Supplementary Information}
% ================================================================================
% 							Fig. S1: Thermal relaxation
% ================================================================================
\noindent \textbf{A. Calculation details} \normalsize \\
\noindent In this section we provide details about the calculations discussed in the main text, and explain how parameter values were determined. First, we consider the steady-state calculations presented in Fig.~1(b) of the main text. We express Eq. 1 of the main text as two coupled ordinary differential equations by defining $w=U \int_0^t ds\,K(t-s)\left(|\alpha(s)|^2-1\right)$. Then
\begin{equation}
    \begin{split}
        i\dot{\alpha}(t) &= \left(-\Delta-i\frac{\Gamma}{2}+w(t) \right)\alpha(t)+i\sqrt{\kappa_1}F\\
        \dot{w}(t) &= \left\lbrace U\left[|\alpha(t)|^2-1\right]-w(t)\right\rbrace/\tau
    \end{split}
\end{equation}
The steady-state solutions are then obtained by requiring $\dot{\alpha}=\dot{w}=0$. The stability of the steady states is assessed by analyzing the spectrum of small fluctuations around the steady-state solution.

For the dynamic simulations based on xSPDE, we use a fourth-order Runge-Kutta algorithm and set a time increment $\Delta t=\Gamma^{-1}/30$. The stochastic terms $\xi_1(t)$ and $\xi_2(t)$  each have zero mean [i.e., $\langle \xi_1 \rangle = \langle \xi_2 \rangle = 0$], and they are delta-correlated with unit variance [i.e., $\langle \xi_1(t) \xi_1(t+t') \rangle=  \langle \xi_2(t) \xi_2(t+t') \rangle =  \delta(t')$]. Moreover,  $\xi_1(t)$ and $\xi_2(t)$  are mutually uncorrelated.

The parameters in Eq.~\ref{eq:IDE} were set to model our experiments as follows. First note that $\Delta$ and $D$ are control parameters, fixed in experiments by the cavity length and the peak-to-peak voltage in the waveform generators driving the modulators. $\Gamma$ is fixed  via measurements of the resonance linewidth at low power. The value of $F$ only matters in relation to the critical amplitude for bistability $F_c$.  As explained in Ref.~\onlinecite{Geng_PRL2020}, by  performing measurements at several powers we identify $(F/F_c)^2 = 1$ as the minimum power for which bistability can be observed. The thermal relaxation time $\tau$, retrieved experimentally as shown in  Supplementary Section B, is far too long to be implemented in our simulations given the memory limitations of our computer. However, this is not a problem because all dynamical effects scale with the ratio $T_\mathrm{mod}/\tau$~\cite{Geng_PRL2020}. Therefore, by fixing this ratio to $T_\mathrm{mod}/\tau=10^3$, as in our experiments, our numerical simulations reproduce our experiments. Finally, the only `free' parameter is $U$. The determination of this parameter is not trivial, since it requires precise knowledge of the intracavity photon number. However, in the absence of quantum effects (as in our experiments), the physics is entirely determined by the ratio  $U|\alpha|^2/\Gamma$. Changing the value of $U$ simply changes the intracavity intensity and the value of $F$ for which bistability is observed, but the spectral lineshape and the PDF remain unchanged provided that $F/F_c$ and $\Delta/\Gamma$ remain unchanged.

The PDFs in Fig.~1(c) and (d) of the main text were constructed as follows. For the selected $F$ and $\Delta/\Gamma$, we first calculated steady-state solutions and assessed their stability. Next we performed $10^4$ stochastic simulations with the value of $\alpha(t=0)$ Gaussian-distributed around the unstable state. The standard deviation of this distribution was $10$\% of the distance in phase space between the stable states. Finally, the PDFs were constructed on basis of all $10^4$ trajectories.\\

\noindent \textbf{B. Thermal relaxation time} \normalsize \\
\begin{figure}[!b]
	\includegraphics[width=\columnwidth]{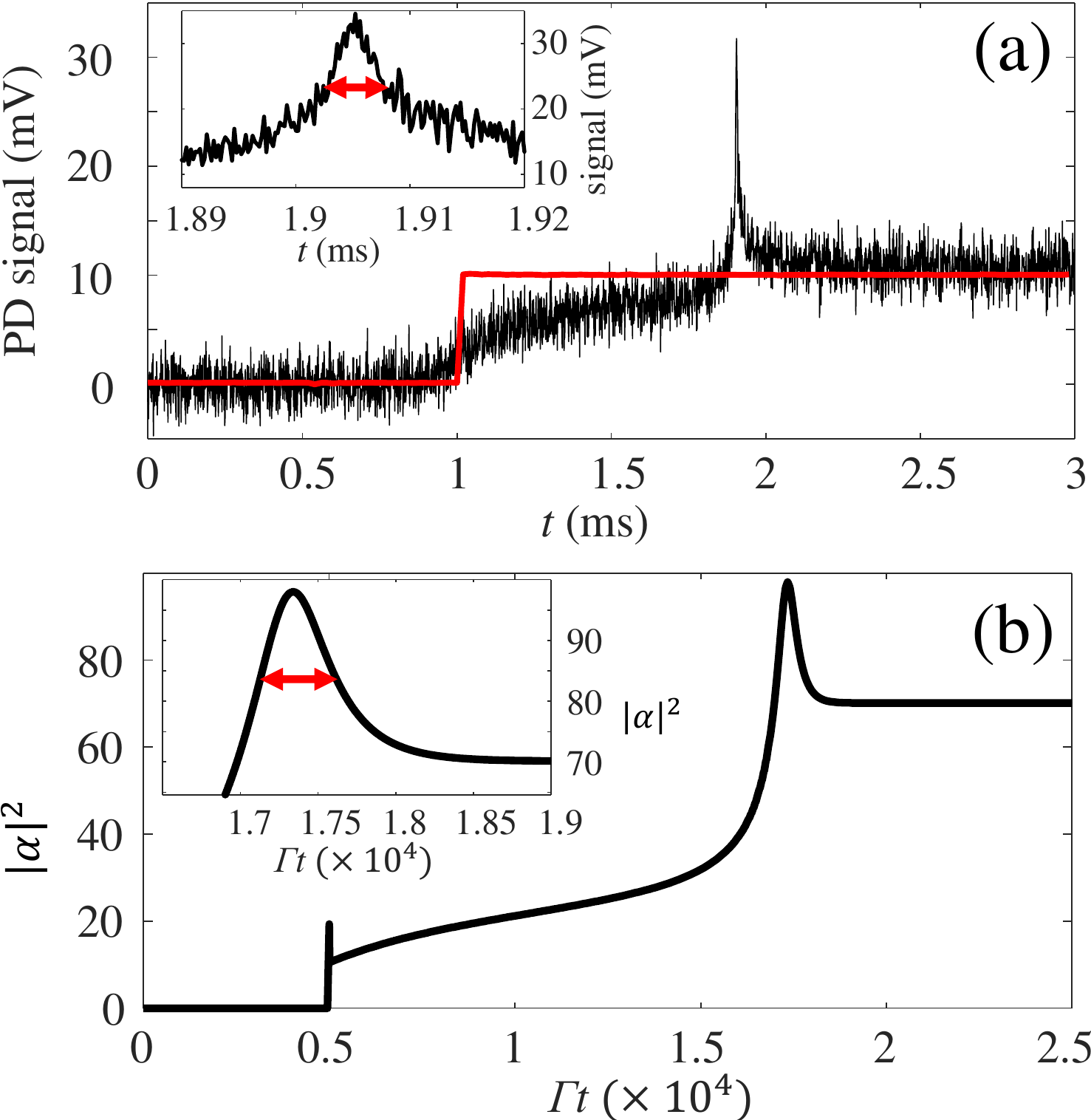}
	\caption{\label{fig:S1}  (a) Intensity transmitted by the cavity (black) when the input laser power (red) is modulated by a chopper. Inset: Zoom into the overshoot. Red arrows indicate the full-width at half-maximum (FWHM), which is $~5$ $\mu$s.  (b) Calculated intensity $|\alpha|^2$ when increasing the driving field from $F=0$ to $F=7\sqrt{\Gamma}$ at $\Gamma t=5000$, thereby driving the system into the upper branch of the bistability. Inset: Zoom into the overshoot. Double-sided arrow indicates the FWHM, corresponding to $\tau/2$.}
\end{figure}
\noindent Here we present a measurement of the thermal relaxation time in our oil-filled microcavity. Our measurement scheme consists of swiftly driving the cavity into the upper branch of the bistability, and subsequently measuring the relaxation time to a steady state. To this end, we measure the cavity transmission while modulating the laser power in a step-like fashion using a chopper; the cavity length is fixed during this measurement. Figure~\ref{fig:S1}(a) shows the chopped input power as a red curve, and the  corresponding cavity transmission as a black curve. When the chopper blocks the laser light, the cavity transmission is negligible. Once the laser passes through an opening in the chopper, the transmission first rises to a transient low-intensity state. Next, there is a build-up phase during which the temperature of the oil, and hence the nonlinearity strength, increases. Once the nonlinearity has built up, the transmission suddenly increases. This leads to a sharp overshoot peak, which is followed by thermal relaxation to a high-intensity steady-state.

In Fig.~\ref{fig:S1}(b) we reproduce the experimentally observed behavior through  numerical simulations based on Eq. 1 in the main manuscript. In the simulations,  we modulate the driving amplitude $F$ in a step-like fashion as done by the chopper in experiments. We performed similar simulations for many values of the model parameters, all in the nonlinear regime. We found that the full width at half maximum (FWHM) of the overshoot is always $\sim \tau/2$ with $\tau$ the memory time of the nonlinearity. Therefore, by analyzing the FWHM of the overshoot in experiments we estimate a memory (thermal relaxation) time $\tau=10$ $\mu$s for our optical microcavity filled with macadamia oil. This value of $\tau$ is close to the one deduced from dynamic hysteresis measurements in a microcavity filled with olive oil, which was found to be $\tau=16$ $\mu$s in Ref.~\onlinecite{Geng_PRL2020}.  \\

% ================================================================================
% 							Fig. S2: Switching events
% ================================================================================
\noindent \textbf{C. Switching behavior} \normalsize \\
\begin{figure}[!b]
	\includegraphics[width=\columnwidth]{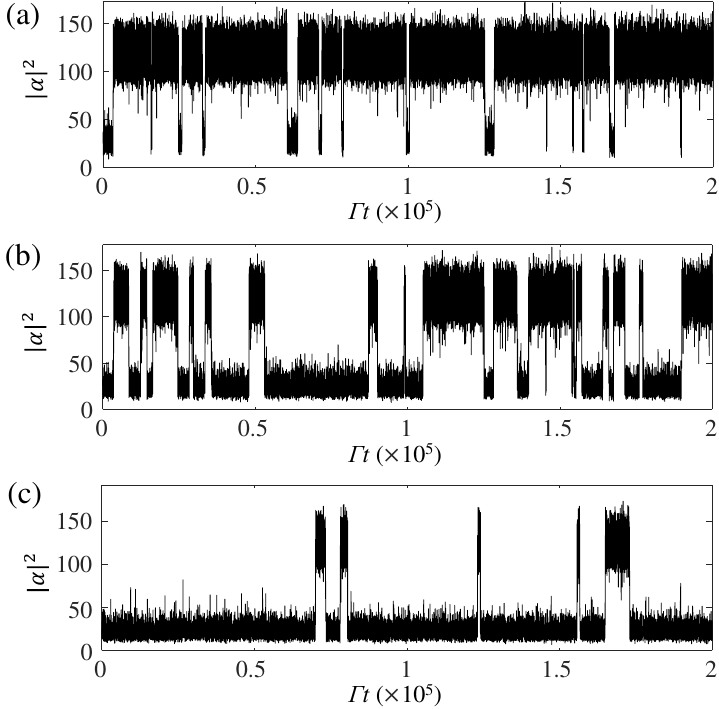}
	\caption{  Stochastic trajectories of $|\alpha|^2$ for different detunings. From (a) to (c) $\Delta/\Gamma=2.4$, $2.5$ and $2.6$, respectively. Model parameters: $F/F_c=2.73$, $\Gamma\tau=4$,  $D=3\Gamma$. }\label{fig:S2}
\end{figure}
\noindent Figure \ref{fig:S2} shows simulated trajectories of $|\alpha|^2$ for constant $F$ and three different $\Delta/\Gamma$, all in the bistability. In all three cases, we observe noise-induced switching between high and low intensity metastable states at random times. The simulations also show how the mean residence times in the metastable states depend on  $\Delta/\Gamma$. The cavity is increasingly biased towards the low intensity state as $\Delta/\Gamma$ increases. This, in turn, changes the shape of the PDFs as shown in Figs.~\ref{fig:1}(c) and (d) of the main text. \\

% ================================================================================
% 							Fig. S3: Noise std vs Vpp
% ================================================================================
\noindent \textbf{D. Noise properties} \normalsize \\
\begin{figure}[!t]
	\includegraphics[width=\columnwidth]{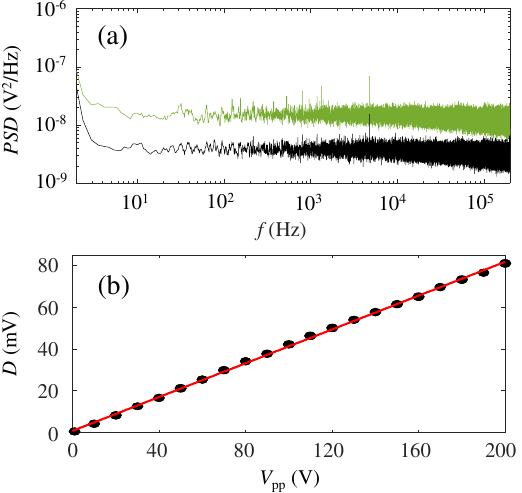}
	\caption{ (a) Measured noise spectrum for two different peak-to-peak voltages $V_\mathrm{pp}$ provided to the modulators. $V_\mathrm{pp}=100$ V for the black curve, and $V_\mathrm{pp}=200$ V for the green curve. (b) Measured standard deviation of the noise $D$ as function of  $V_\mathrm{pp}$. The red line is a linear fit to the data.}\label{fig:S3}
\end{figure}
\noindent Here we present a characterization of the noise imprinted on the driving laser by the electro-optic modulators (EOMs), to which we supply a peak-to-peak voltage $V_\mathrm{pp}$. To this end, we measured the laser power directly after the second EOM while adding noise of variable strength. Figure~\ref{fig:S3}(a) shows the noise spectrum for two different $V_\mathrm{pp}$. The measurement time was $50$ seconds, and the sampling rate was $1$ MS/s. We observe an approximately flat spectrum corresponding to white noise, and a greater noise floor for greater $V_\mathrm{pp}$. Figure~\ref{fig:S3}(b) shows the  standard deviation of the noise as a function of the $V_\mathrm{pp}$ provided to the EOMs by the waveform generators plus amplifiers. The measurement time was $1$ second, and the sampling rate was 10 MS/s. We observe a linear increase of the standard deviation of the noise with $V_\mathrm{pp}$.  \\

\begin{figure}[!t]
	\includegraphics[width=\columnwidth]{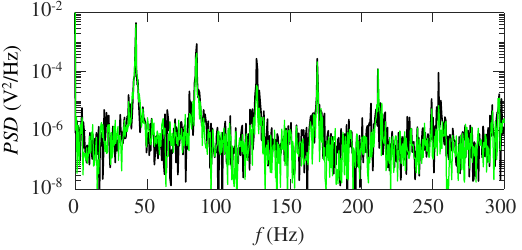}
	\caption{  Measured power spectral density of the transmitted laser while modulating the cavity length. The peak-to-peak voltage provided to the modulators is $V_\mathrm{pp}=70$ V for the black curve and $V_\mathrm{pp}=200$ V for the green curve. Experimental settings for the black and green curve are the same as in Fig. 2(g) and 2(h) of the main text, respectively.}\label{fig:S4}
\end{figure}

% ================================================================================
% 			   Fig. S4: PSDs wide freq. range
% ================================================================================
\noindent \textbf{E. Decrease in output power} \normalsize \\
\noindent Here we explain why the $SNR$ corresponding to the trajectory in Fig.~2(d) of the main text is lower than the $SNR$ corresponding to the trajectory in Fig.~2(c). In Fig. \ref{fig:S4} we plot the power spectral density $PSD$ corresponding to the  measurements in Figs.~2(c) and (d). The $PSD$s show peaks corresponding to the first six harmonics. The height of these peaks is lower for the transmitted signal at $V_\mathrm{pp}=200$ V, most notably for the third and sixth harmonics. This implies a lower total signal power at $V_\mathrm{pp}=200$ V. Hence,  although Figs. 2(c) and (d), show  a similar noise floor for $V_\mathrm{pp}=70$ V and $V_\mathrm{pp}=200$ V, the $SNR$ is lower for $V_{PP}=200$ V.\\

% ================================================================================
% 							Fig. S5: Changepoint detection
% ================================================================================
\noindent \textbf{F. Detection of switching events} \normalsize \\
\noindent To compute the average number of switching events per modulation cycle $n_\mathrm{avg}$, we used the standard Matlab function `findchangepts'. Given an $N$-point time series $\left\lbrace x_1,...,x_N \right\rbrace$, the function determines the $K$ change points $\left\lbrace x_k\right\rbrace$ minimizing the cost function
\begin{equation}\label{cost}
	J(K)=\sum_{r=0}^{K-1}\sum_{i=k_r}^{k_{r+1}-1}\delta\left[x_i;\chi\left(\left\lbrace x_{k_r},...,x_{k_{r+1}-1}\right\rbrace\right)\right]+\beta K.
\end{equation}
Here, $k_r$ is the index of the $r^{th}$ change point. $\delta$ gives the difference between a point $x_i\in \left\lbrace x_{k_r},...,x_{k_{r+1}-1}\right\rbrace$ and the mean $\chi$ of the subtrajectory $\left\lbrace x_{k_r},...,x_{k_{r+1}-1}\right\rbrace$. $\beta$ is a fixed penalty which is added for each change point.

\begin{figure*}[!]
	\includegraphics[width=\textwidth]{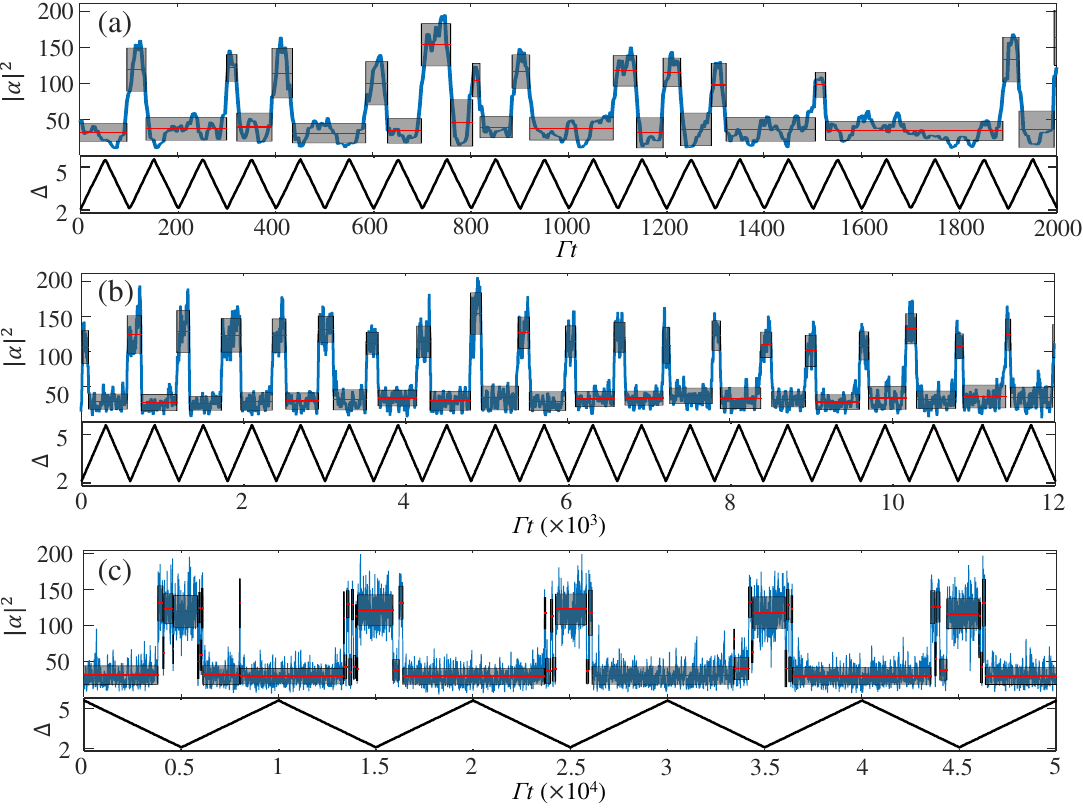}
	\caption{\label{fig:S5}  Simulated intensity (blue lines) as a function of time (multiplied by the total loss rate $\Gamma$) for three different modulation periods (black lines), at fixed noise variance. Subtrajectories following from the changepoint detection algorithm are indicated by their mean (red lines) and standard deviation (shaded box).  From (a) to (c) $\Gamma T_\mathrm{mod}=100$, $\Gamma T_\mathrm{mod}=600$, $\Gamma T_\mathrm{mod}=10^4$. Simulation parameters are as in Fig.~\ref{fig:4} of the main text.}
\end{figure*}

The number of detected change points generally depends on the value of $\beta$. If $\beta$ is too small (large), the algorithm returns too many (few) change points. Therefore, the right value of $\beta$ needs to be determined for each data set. To avoid false or missing change points related to the value of $\beta$, we implemented the following algorithm. First we set a relatively small value of $\beta$ and get all true change points and extra ones. Next, we compute the mean and standard deviation of each subtrajectory. If the difference in the mean of two consecutive subtrajectories is less than the sum of their standard deviations, we remove the change point that separates the two subtrajectories. This process is repeated until no change points are removed anymore.

Figure~\ref{fig:S5} illustrates the analysis of three representative trajectories using our algorithm. The mean and standard deviation of each subtrajectory is indicated by a red line and shaded box, respectively. Figures~\ref{fig:S5}(a), ~\ref{fig:S5}(b), and ~\ref{fig:S4}(c) show trajectories for which $n_\mathrm{avg}<2$, $n_\mathrm{avg}\sim 2$ and  $n_\mathrm{avg}>2$, respectively. In Fig.~\ref{fig:S5}(c) we show fewer modulation periods to make the multiple switching events per cycle visible in the plot. Overall, the results in Fig.~\ref{fig:S5} demonstrate that our algorithm successfully detects all true change points and does not give any false change points. \\

\begin{figure}[!b]
	\includegraphics[width=\columnwidth]{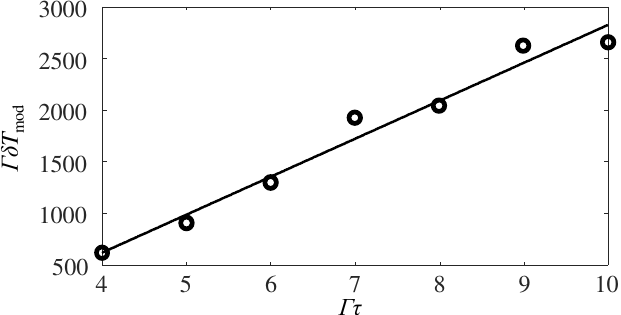}
	\caption{\label{fig:S6} $\delta T_\mathrm{mod}$ is the range of modulation periods for which the average number of switches per cycle is $n_\mathrm{avg}\in \left[2-\epsilon,\, 2+\epsilon\right]$, with $\epsilon=0.1$. $\delta T_\mathrm{mod}$ (multiplied by the total loss rate $\Gamma$)  is shown as a function of the memory time $\tau$.   Model parameters are the same as in Fig. \ref{fig:4} of the main text.}
\end{figure}

% ================================================================================
% 							Fig. S6: Scaling of frequency range SR
% ================================================================================
\noindent \textbf{G. Memory-enhanced stochastic resonance bandwidth} \normalsize \\
\noindent  Figure~\ref{fig:4} in the main text shows the emergence of a `plateau' at $n_\mathrm{avg}\approx2$ as $\tau$ increases. That `plateau'  represents an enlargement of the frequency range in which stochastic resonance (SR) can be achieved, i.e., the SR bandwidth. Here we show that the SR bandwidth  increases linearly with $\tau$. To this end, in  Fig. \ref{fig:S6} we plot the range of modulation periods $\delta T_\mathrm{mod}$ for which $n_\mathrm{avg}\in \left[2-\epsilon,\, 2+\epsilon\right]$. Results of numerical simulations (circles) are fitted by a line, demonstrating that $\delta T_\mathrm{mod}$ increases linearly with the memory time $\tau$. We confirm that the relation between $\delta T_\mathrm{mod}$ and $\tau$ remains linear for any $\epsilon \ll 2$.

\end{document}